\documentclass{aa}
\input{psfig.sty}
\usepackage{graphics}
\usepackage{natbib}
\usepackage{version}
\usepackage{txfonts}
\bibpunct{(}{)}{;}{a}{}{,} 

\newcommand{\noun}[1]{\textsc{#1}}

\def\ion#1#2{#1{\sc#2}}
\def\HI{\ion{H}{i}}
\def\lya{Ly-$\alpha$}
\def\Horizon{\mbox{\noun{Horizon}-$4\Pi$}}

\def\Sec#1{Section~\ref{sec:#1}}

\begin{document}

  \title{Simulations of BAO reconstruction
  with a quasar \lya\ survey}
 \titlerunning{simulations of quasar \lya\ survey}

 \author{J.M. Le Goff\inst{1}\thanks{\email{jmlegoff@cea.fr}} 
 \and C. Magneville \inst{1}
 \and E. Rollinde \inst{2} 
 \and S. Peirani \inst{2} 
 \and P. Petitjean \inst{2} 
 \and C. Pichon \inst{2} 
 \and J. Rich \inst{1}
 \and C. Yeche \inst{1}
 \and E. Aubourg \inst{3,1}
 \and N.~Busca \inst{3}
 \and R. Charlassier \inst{1}
 \and T. Delubac \inst{1}
 \and J.C. Hamilton \inst{3}
 \and N. Palanque Delabrouille \inst{1} 
 \and I. P\^aris \inst{2} 
 \and M.~Vargas \inst{3}
 }
 \institute{CEA  centre de Saclay, irfu/SPP, F-91191 Gif-sur-Yvette, France  
 \and Institut d'Astrophysique de Paris, UMR7095 CNRS, Universit\'e Pierre et
Marie Curie, 98 bis bd Arago, 75014 Paris, France
 \and APC, 10 rue Alice Domon et L\'eonie Duquet, F-75205 Paris Cedex 13, France }

 \authorrunning{J.M. Le Goff et al.}

\date{Received xx xx 2011 / accepted  xx xx 2011}


   \abstract
{The imprint of Baryonic Acoustic Oscillations (BAO) on the matter power spectrum can be constrained using the neutral hydrogen density in the intergalactic medium (IGM) as a tracer of the matter density. One of the goals of the Baryon Oscillation Spectroscopic Survey (BOSS) of the Sloan Digital Sky Survey (SDSS-III) is to derive the Hubble expansion rate and the angular scale from the BAO signal in the IGM. To this aim, the Lyman-$\alpha$ forest of $10^5$ quasars will be observed in the redshift range $2.2<z<3.5$ and over $\sim 10,000$ deg$^2$.
}
{We simulated the BOSS QSO survey to estimate the statistical accuracy on the BAO scale determination provided by such a large scale survey. 
 In particular, we discuss the effect of the poorly constrained estimate of the unabsorbed intrinsic quasar spectrum.
}
{ The volume of current $N$-body simulations being too small for such studies, we resorted to Gaussian random field (GRF) simulations. We validated the use of GRFs by comparing the output of GRF simulations with that of the \Horizon\ $N$-body dark-matter-only simulation with the same initial conditions. Realistic mock samples of QSO Lyman-$\alpha$ forest were generated; the3ir power spectrum was computed and fitted to obtain the BAO scale. The rms of the results for 100 different simulations provides an estimate of the statistical error expected from the BOSS survey.}
{ We confirm the results from Fisher matrix estimate. In the absence of error on the unabsorbed quasar spectrum, the BOSS quasar survey should measure the BAO scale with an error of the order of 2.3\%, or the transverse and radial BAO scales separately with errors of the order of 6.8\% and 3.9\%, respectively. The significance of the BAO detection is assessed by an average $\Delta\chi^2=17$ but for individual realizations $\Delta\chi^2$ ranges from 2 t o 35. The error on the unabsorbed quasar spectrum increases the error on the BAO scale by 10 to 20\% and results in a sub percent bias.
}
{}


\maketitle

\section{Introduction}

Constraining the properties of dark energy that drives the expansion of the Universe is key towards understanding cosmology. Baryonic acoustic oscillations (BAO) in the baryon-photon fluid of the pre-recombination Universe imprint the sound horizon distance at decoupling  as  a typical scale in the matter correlation function or power spectrum \citep{peeblesal70,sunyaeval70,eisensteinal98,bashinskyal02}. These oscillations were detected both in the cosmic microwave background \citep[e.g.][]{pageal03} and in the spatial distribution of galaxies at low redshift \citep{eisensteinal05,coleal05,percivalal09}. Their measurements give important and coherent constraints on cosmological parameters \citep{komatsual09}.

More recently, it was realized that BAOs could be detected in the \lya\ forest { \citep{mcdo03,2003dmci.confE..18W}} used as a probe of the intergalactic medium (IGM) at intermediate redshifts ($z\sim 2-3$) { and the potential of the measurement was quantified by \citet{mcdoal07}}. The  structure and composition of the IGM has long been studied using the \lya\ forest in QSO absorption spectra \citep{rauch98}. The advent of high spectral resolution Echelle-spectrographs on 10~m-class telescopes has led to a consistent picture in which the absorption features are related to the distribution of neutral  hydrogen (\HI) through the \HI\ Lyman transition lines. The IGM is believed to contain the majority of baryons in the Universe at these redshifts \citep[]{petitjeanal93,fukugitaal98}, and is highly ionized by the UV-background produced by galaxies and QSOs \citep[]{gunnal65}, at least since $z\sim 6$ \citep[]{fanal06,beckeral07}. Photo-ionization equilibrium in the expanding IGM establishes a tight correlation between neutral and total hydrogen densities. Numerical simulations and analytical models support the existence  of this correlation and show that the gas density traces the  fluctuations of the DM density on scales larger than the Jeans length \citep[see for example][]{bial92,cenal94,petitjeanal95,miraldaal96,theunsal98}.

In this paradigm, the IGM consists of mildly non-linear gas that traces the dark matter, and is photo-heated by the UV-background. Although metals are detected in the IGM \citep[]{Cowie95,Schaye03,aracilal04}, stirring of the IGM due to feedback from galaxies and AGNs probably does not strongly affect the vast majority of the baryons \citep[e.g.][]{Theuns02b,Mcdonald05}. 
{ The relation between the \lya\ forest flux and the underlying matter field is non-linear since fluctuations are compressed to the range $0<F<1$.  However, unlike galaxy surveys which sample only peaks  in the matter field, the whole space along the quasar line-of-sight is democratically sampled.  This is expected to lead to less scale-dependence in the bias compared to that observed in galaxy surveys.}
The shapes and clustering of lines have been extensively used to infer the temperature of the IGM \citep[]{Schaye99, Ricotti00,Theuns00, McDonald01}, determine the amplitude of the UV-background \citep[]{Rauch97, Bolton05}, trace the density structures around galaxies and quasars \citep[]{Rollinde,Guimaraes,KimCroft}, constrain the reionization history of the Universe \citep[]{Theunsal02a, Hui03, fanal06}, measure the matter power spectrum \citep[]{Croftal99,Viel04,McDonald06} or constrain cosmological parameters  \citep[]{McDonald99,Rollinde03,Coppolani,Guimaraes,Viel06}. \cite{paddyal09} \citep[see also][]{meiksinal99} analyzed the amplitude of non-linear effects by comparing perturbative theory with outputs of ten dark matter numerical simulations. Although they focused on halos only, they demonstrated that the shift of the reconstructed BAO scale due to non-linearities decreases with redshift as $D^2(z)$, the square of the linear growth factor. The simple linear bias between flux and matter power spectrum was also predicted by \citet{mcdo03} at low $k$-values, and by \citet{slosaral09} and \citet{whiteal09} at BAO scales.

The observation of BAOs in the \lya\ forest requires a full 3-dimensional sampling of the matter density, and therefore a much higher number density and number of quasars than previously available. The Baryon Oscillation Spectroscopic Survey (BOSS) \citep{schlegelal09} of the Sloan Digital Sky Survey-III (SDSS-III) \citep{Eisenstein11} aims to identify and observe more than 150,000 QSOs over 10,000 square degrees. The QSO redshift range useful for BAO reconstruction is limited to $z>2.15$ on the low side by the requirement that the \lya\ absorption falls in BOSS spectrograph wavelength range. It is limited  to about $z<3.5$ on the high side by the sharp decrease of the QSO density both intrinsically and for the magnitude accessible to the spectrograph ($g<22$). \cite{mcdoal07} estimated, from standard Fisher matrix techniques and analytical description of the \lya\ power spectrum, that the   density of quasars should   be of the order of 20-30 per square degree to achieve  constraints of the order of 1\%\ in the radial and transverse BAO scales, see also \cite{McQuinnWhite11}. Such a high requirement lead to new developments on  target selection \citep{NPDal10,yecheal10,Bovyal11,Kirkpatrick11,Rossal11}.

It is  important to confirm the predictions on the BAO scale measurements with additional work on numerical simulations. Recently, \cite{slosaral09,whiteal09} have studied the BAO signature in a typical  \lya\ forest survey, using large $N$-body simulations, but with still higher quasar density. This paper follows those works and investigates errors in the BAO scale estimates as a function of the properties of the survey such as the density of quasars and the amplitude of the noise.  Very recently \cite{GreigBolton11} have published a similar study but assuming all QSO located at the same $z$, with constant $S/N$ ratio and a much smaller volume than the BOSS survey (corresponding to 79 instead of 10,000 deg$^2$).
In Sect.~\ref{sec:simulations} a comparison with the \Horizon \citep{prunetal08} dark-matter-only $N$-body simulation validates the use of linear Gaussian random field (GRF) to study BAO scale reconstruction. The production of realistic mock spectra is described, including quasar unabsorbed spectrum, noise, and the effect of peculiar velocities. Standard methods to analyze BAO signal in terms of power spectrum are presented in \Sec{methods}. The performance of the survey for different quasar densities and noise amplitudes are presented in Sect.~\ref{sec:results} and the effect of the error on the estimate of the quasar unabsorbed spectrum is discussed. The resulting cosmological constraints are presented in Sect.~\ref{sec:cosmo} and we draw  conclusions in Sect.~\ref{sec:discussion}.

\section{Description of the simulations}
\label{sec:simulations}

\subsection{From Nbody to Gaussian random field (GRF) simulations}
\label{sec:Horizon}

The size of large N-body simulations is typically (2 Gpc/$h$)$^3$. At $z=2.5$ this corresponds to 745 deg$^2$ which is much smaller than the 10,000 deg$^2$ of the BOSS survey. As will be clear in Sect.~\ref{sec:resuTrueC}, probing such a volume with the \lya\ forest of 20 quasars per deg$^2$ results in a power spectrum or a correlation function where the BAO features are hardly seen: some realizations will exhibit them and some will not. A larger volume is therefore required in order to study the error on the reconstructed BAO scale. 

Such a volume can be provided by Gaussian random field simulations, which however do not contain any non-linear effects.The \noun{Horizon}-$4\Pi$ simulation was used to investigate the relevance of these effects for the study of BAO scale reconstruction. \noun{Horizon}-$4\Pi$ is  a $\Lambda$CDM dark-matter-only simulation based on cosmological parameters inferred by the WMAP three-year results, with a box size of $2h^{-1}$Gpc on a grid of size $4096^{3}$. The purpose of this simulation is to investigate full sky weak lensing and baryonic acoustic oscillations. The 70 billion particles were evolved using the Particle Mesh scheme of the \emph{\noun{RAMSES}} code 
on an adaptively refined grid (AMR) with about 140 billion cells. Each of the 70 billions cells of the base grid was recursively refined up to 6 additional levels of refinement, reaching a formal resolution of 262,144 cells in each direction (roughly 7 kpc/$h$ comoving). The code \texttt{mpgrafic} \citep{prunetal08} was used to generate the initial conditions (ICs). 

For identical initial conditions, we compared outputs from \Horizon\ simulation with the linear density modified  through a lognormal model 
to incorporate some of the non-linearities. To make this comparison statistically as powerful as possible we did not implement here all the complications of a realistic survey. Transmitted-flux-fraction spectra were generated without any observational noise and extended in wavelength all along the box, instead of being limited to the \lya\ forest. In addition the $x$ and $y$ positions of those lines were chosen regularly which eliminates the sampling noise (see Sect.~\ref{sec:GRF} and Eq.~\ref{Eq:Pk}). In this case, we can divide the simulation into eight individual boxes with the same volume and still see the BAO features in each box. The FFT of each box was computed to produce the power spectrum, which was Fourier transformed to give the correlation function. The correlation function was fitted to determine the BAO peak position. This was done both for the full \Horizon\ and for the lognormal densities, yielding $k_A = 0.05766 \pm 0.00197$ h/Mpc (LN) and $k_A = 0.05820 \pm 0.00192$ h/Mpc (Horizon), where the errors are obtained from the rms of the 8 values. The two sets of values are correlated and the difference is $0.00053 \pm 0.00166$, i.e.~($0.9\pm 2.9$)\%. We conclude that we do not observe non-linear effect on the reconstruction of the BAO scale at this level of accuracy.

\subsection{Gaussian random field simulations}
\label{sec:GRF}

Complex normal Gaussian fields were generated in a box in Fourier space with $\delta(-k) = \delta^*(k)$. The amplitude of each mode was multiplied by the square root of the power spectrum $P(k)$ at $z=0$ from \citet{eisensteinal98} with $H_0=71$ km/s/Mpc, $\Omega_m=0.27$, $\Omega_b=0.044$, $\Omega_\Lambda=0.73$ and $w=-1$. An inverse FFT provided a linear simulation of matter density fluctuations, $\delta \rho/\rho$, at $z=0$ in a box of $2560 \times 2560 \times 512$ pixels of (3.2 Mpc/$h)^3$, i.e.~8179 Mpc/$h$ in the transverse directions and 1636 Mpc/$h$ in the longitudinal direction. The transverse size covers 12,500 deg$^2$ at $z=2.5$ and 9100 deg$^2$ at $z=3.5$, while the longitudinal size covers from z=1.75 (corresponding to $\lambda=3344\; \AA $ for \lya\ absorption) to $z=3.9$. 
Peculiar velocities in the direction parallel to the line of sights were computed from densities in Fourier space using $\displaystyle v_{k_\parallel}=-\frac{ik_\parallel}{k^2}\frac{H(z)}{D(z)}\frac{\partial D}{\partial z}(z) \delta_k$ where $D(z)$ is the linear growth factor.

The clustering of quasars was neglected and their angular positions were randomly drawn within the box. They were assigned a redshift and a magnitude according to the distribution of \citet{jiangal06}. The lines of sight of the quasars were taken to be parallel. Real data will have to be analyzed taking into account the angle between the lines of sight, but if we consistently produce the simulation and analyze the resulting spectra with parallel lines of sight, this should have a negligible effect on the statistical error on the reconstructed BAO scale.

\begin{figure} 
\resizebox{\hsize}{!}{
\includegraphics{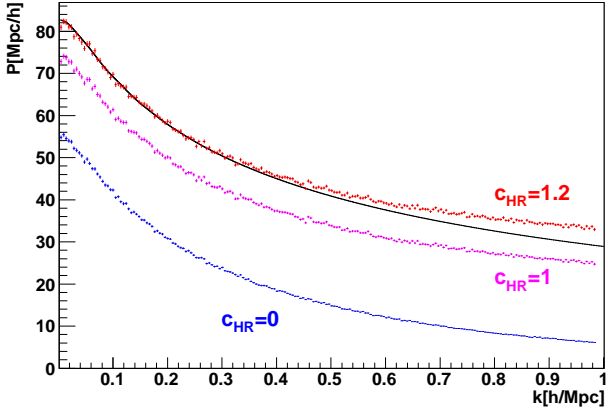}} 
\caption{One-dimension power spectrum ($P^{1D}$) for our simulation with $c_{HR}=0$ (lower, blue histogram), $c_{HR}=1$ (magenta) and $c_{HR}=1.2$ (upper, red), where $c_{HR}$ is defined in Eq.~\ref{Eq:LRHR}, compared to expectation from Eq.~\ref{Eq:KP} (black continuous line).} 
\label{fig:P1D}
\end{figure}

The matter density, in pixels located along the line of sight of each quasar, was evolved to the redshift of the pixel by multiplying $\delta \rho / \rho$ by $D(z)$ at the redshift of the considered pixel. This provided us with density spectra in 3.2 Mpc/$h$ bins. We did not interpolate to regular bins, e.g.~in $\log(\lambda)$, and then interpolate back to regular bins in Mpc to compute the power spectrum. Since we anyway average the flux over (12.8 Mpc/$h)^3$ pixels before applying a FFT to compute the power spectrum (Sect.~\ref{sec:Pk}), we believe this is a minor effect.

Because QSOs sample only a very small transverse region, such a simulation misses a significant contribution to the power spectrum coming from transverse scales smaller than the 3.2 Mpc/$h$ pixel size. This is very clear for the 1D power spectrum, which is an integral of the 3D power spectrum over $k_\perp$ \citep{KaiserPeacock91} and therefore includes large $k_\perp$ :
\begin{equation}
\label{Eq:KP}
P^{1D} (k_\parallel) = \frac{1}{2\pi} \int_0^\infty  P(k_\parallel,k_\perp) k_\perp dk_\perp.
\end{equation}
This missing small scale contribution is illustrated in Fig.~\ref{fig:P1D}, where the 1D power spectrum of the matter density for the simulation with 3.2 Mpc/$h$ pixels (lower blue histogram, $C_{HR}=0$) appears significantly lower than the expectation from Eq.~\ref{Eq:KP} (black continuous curve). In order to compensate for the missing contribution, we have generated 20 high-resolution (HR) Gaussian random field simulations with a volume corresponding to one pixel of our large-volume low-resolution simulations (LR), i.e.~(3.2 Mpc/$h)^3$, and a pixel size of 200 kpc. When simulating the matter density along a quasar line-of-sight, for each large pixel we randomly selected one of the HR simulations, and we defined the density in the small pixels as 
\begin{equation}
\label{Eq:LRHR}
\delta \rho = \left( \delta \rho \right)_{LR} + c_{HR} \left( \delta \rho \right)_{HR}.
\end{equation}
As illustrated in Fig.~\ref{fig:P1D}, if we just add the LR and HR simulations (i.e.~$c_{HR}=1$ in Eq.~\ref{Eq:LRHR}) the 1D power spectrum of the matter density is still slightly smaller than predicted by Eq.~\ref{Eq:KP} 
\footnote{  Note that in Eq.~\ref{Eq:KP} we integrated up to $k_\perp=\pi/$(100 kpc/$h$) which corresponds to a typical value of the Jeans' scale. This however is only a reduction of a few \% relative to the integral up to infinity. 
}.
This is not surprising since we do not have any correlation between the HR simulations in neighboring large pixels. Using an effective correction factor $c_{HR}=1.2$ results in a $P^{1D}(k)$ which fits well Eq.~\ref{Eq:KP}, at least in the $k_\parallel$ range relevant for BAO, see Fig.~\ref{fig:P1D}. 


 \citet{mcdoal07} showed that the observed power spectrum, $P_{\rm obs}(\vec k)$, is the sum of the true power spectrum, a sampling contribution and a noise contribution :
\begin{equation}
\label{Eq:Pk}
  P_{\rm obs}(\vec k) = P(\vec k) + P_W^{2D} P^{1D}(k_\parallel) + P_N, 
\end{equation}
where, in the absence of pixel weighting, $P_W^{2D}$ is the inverse of the surface density of quasars (in Mpc$^{-2}$). Eq.~\ref{Eq:Pk} indicates that the accurate description achieved for $P^{1D}$ at $k_\parallel \le 0.3$, ensures a good description of $ P_{\rm obs}(\vec k)$ in the range relevant for BAO.

The next step was to go from the matter density fluctuations, to QSO transmitted flux fractions $F=\exp( -\tau)$, where $\tau$ is the optical depth. Note that $F$ is the traditional notation for the transmitted flux fraction and we will use $\phi$ for the QSO flux. We used the relation
\begin{equation}
\label{Eq:FGPA}
F =\exp \left[- a(z) \exp b \frac{\delta \rho}{\rho} \right]. 
\end{equation}
This means the lognormal approach was used to get the baryon density from our Gaussian fields \citep{BiDavidsen1991} and the baryon density was transformed into transmitted flux fractions $F$ using the fluctuating Gunn-Peterson approximation  
\citep{Croftal98,1998MNRAS.296...44G}.  

We followed the procedure of \citet{mcdo03} to take into account the effect of the peculiar velocities: we accounted for the expansion or contraction of cells by translating each cell edge in real space into redshift space using the average velocity of the two cells that the edge separates. The optical depth contributed by each real-space cell was then distributed to multiple redshift-space pixels based on its fractional overlap with each.

\begin{figure} 
\resizebox{\hsize}{!}{
\includegraphics{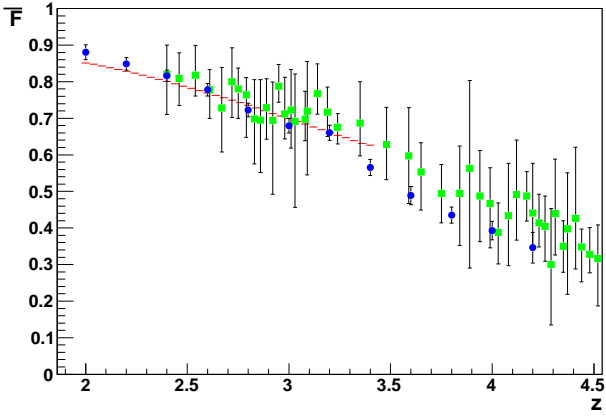}} 
\caption{Mean transmitted flux fraction, $\overline{F}(z)$, as a function of redshift for the simulation (red histogram) compared to \citet{FG2008} data (blue circles) and  \citet{2004AJ....127.2598S} data (green squares).} 
\label{fig:F}
\end{figure}

The value of $b$ in Eq.~\ref{Eq:FGPA} was fixed to $b=2-0.7(\gamma -1)=1.58$ for an equation of state parameter $\gamma=1.6$
\citep{1997MNRAS.292...27H}. 
The value of $a(z)$ was fitted to reproduce the experimental 1D power spectrum and  the resulting mean transmitted flux fraction $\overline{F}(z)$ was checked to be in good agreement with the data, as illustrated by Fig.~\ref{fig:F}. We could alternatively have fitted $\overline{F}(z)$ and checked $P^{1D}$. More precisely, \citet{McDonald06} measured $P^{1D}$ for $k_\parallel$ between $ \approx 0.14$ and $\approx 1.8 h/$Mpc and $2.2<z<4.2$, and in each bin in $z$ we fixed $a(z)$ to fit the first four bins in $k$, from $\approx$0.14 to $\approx$0.28 $h$/Mpc. We did not fit higher $k$ bins because our simulations do not include non-linear effects and are not expected to fit data at high $k$. 
Note that it would have been more natural to use 100 kpc/$h$ pixel size for the HR simulations, a typical value of the Jeans scale. In this case one needs a correction factor $c_{HR}$=1.12 only, but one cannot simultaneously reproduce $\overline{F}(z)$ and $P^{1D}$.

\begin{figure} 
\resizebox{\hsize}{!}{
\includegraphics{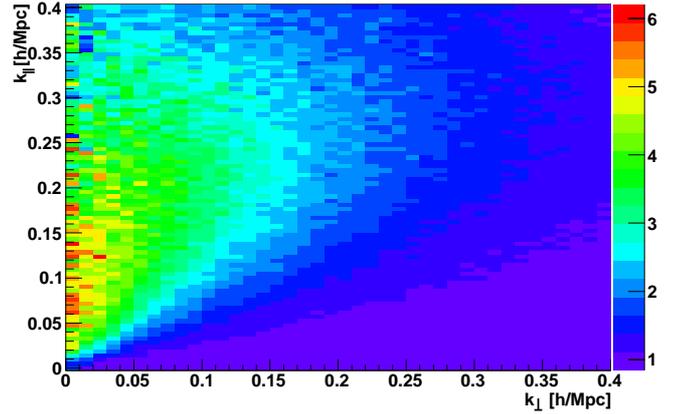}} 
\caption{Two-dimension power spectrum $P(k_\perp,k_\parallel)$ in redshift space
divided by the 2D power spectrum in real space.} 
\label{fig:P2D}
\end{figure}

Peculiar velocities introduce a dependence on $\mu=k_\parallel/k$ in the redshift-space power spectrum :  on large scales we have $ P(k,\mu) = (1+\beta \mu^2)^2 P_0(k)$, where $P_0(k)$ is the isotropic real-space power spectrum \citep{1987MNRAS.227....1K}. 
For galaxy surveys, $\beta$ is related to the bias and the growth rate of structure, but for \lya\ forest it is an independent parameter \citep{2000ApJ...543....1M}. 
Fig.~\ref{fig:P2D} shows the ratio of the redshift-space over the real-space power spectra for our simulation\footnote{Note that the power spectra were obtained without noise and using all pixels of the box, not just those along some random QSO lines. This removes the contribution from the noise and sampling terms so that $P_{\rm obs}(k) = P_F(k)$, see Eq.~\ref{Eq:Pk}.}. This ratio follows Kaiser formula in the $k$ range relevant for BAO, $k<0.2$. The departure at higher $k$ is due to the fact that our procedure to implement the effect of velocities is only valid for scales larger than a few ($3.2$Mpc$/h$) pixels. The ratio of power spectra is unity in the transverse direction ($\mu=0$) and about 5 in the longitudinal direction ($\mu=1$), which corresponds to  $\beta=1.2$ in Kaiser formula. This is to be compared to $\beta=1.58$ according to \citet{mcdo03} simulations and $0.38<\beta<1.05$, as measured with first BOSS data~\citep{xipush}. { Note, however, that the value of $\beta$ obtained by BOSS is contaminated by the presence of damped \lya\ systems and metal lines in the quasar spectra}.
 We also observe a bias $b=0.19$ relative to the matter power spectrum, to be compared to $0.17<b<0.25$ measured with BOSS data.

\begin{figure} 
\resizebox{\hsize}{!}{
\includegraphics{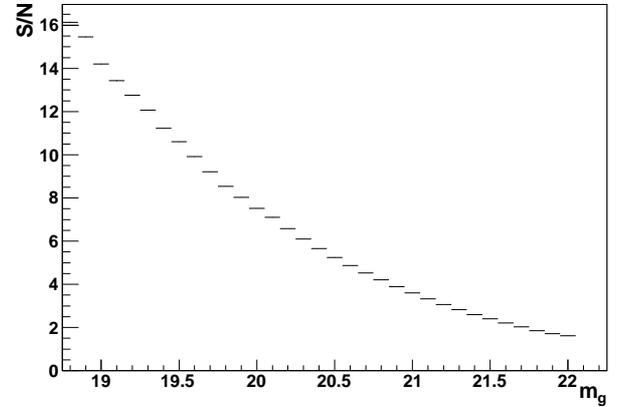}} 
\caption{Mean signal-to-noise ratio in the \lya\ forest for 1\AA\ bins, as a function of the quasar magnitude in the g band.} 
\label{fig:sn}
\end{figure}

\begin{figure} 
\resizebox{\hsize}{!}{
\includegraphics{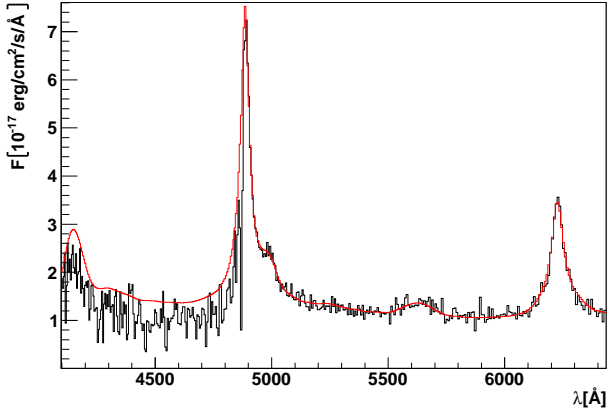}} 
\caption{Mock spectrum of a quasar with redshift $z=3.02$ and magnitude $m_g=21.30$. The continuous red line is the input PCA unabsorbed spectrum.} 
\label{fig:spectrum}
\end{figure}

\begin{figure} 
\resizebox{\hsize}{!}{
\includegraphics{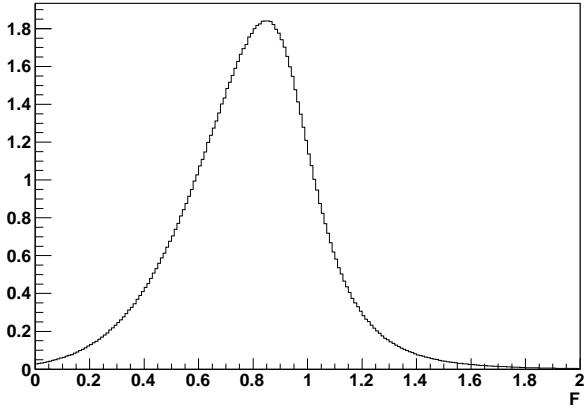}} 
\caption{Probability distribution function of the transmitted flux fraction in the 3.8 Mpc/$h$ simulation bins. }
\label{fig:pdf}
\end{figure}

To get the flux, $\phi_i(\lambda)$, of quasar $i$, the transmitted flux fraction, $F_i(\lambda)$, must be multiplied by the quasar unabsorbed spectrum, i.e.~the quasar spectrum, including the QSO emission lines, if there were no absorption. The principal component analysis (PCA) of \citet{2005ApJ...618..592S} 
was used to generate for each mock spectrum a random PCA unabsorbed spectrum, which was normalized according to the g-band magnitude of the quasar. Noise was added according to the characteristics of BOSS spectrograph, including readout noise, sky noise and signal noise and assuming four exposures of 900 s, for each QSO spectrum. 
Fig.~\ref{fig:sn} presents the mean signal-to-noise ratio per 1$\AA$ bin in the \lya\ forest, which varies from 14 for a quasar magnitude $m_g=19$ to 1.6 for $m_g=22$.
Fig.~\ref{fig:spectrum} shows an example of such a mock spectrum.
The pdf of the transmitted flux fraction is presented in Fig.~\ref{fig:pdf}. For high resolution bins this pdf exhibits peaks at zero and unity but for low resolutions bins, the flux is averaged and there is a single peak around $\overline F$. In addition, due to noise the transmitted flux fraction can be larger than unity and also negative.

\section{From spectra to BAO signal }
\label{sec:methods}

The mock spectra, produced as described above, were analyzed to reconstruct the power spectrum and the BAO scale.

\subsection{Power spectrum estimator}
\label{sec:Pk}

The flux $\phi_i$ in mock spectrum $i$ must be divided by the product of $ \overline{F}(z)$  and the quasar unabsorbed spectrum $c_i$ in order to provide the fluctuation of the transmitted flux fraction, $\delta_F (\lambda)$, as
\begin{equation}
\label{Eq:delta_F}
\delta_F(\lambda) = \frac{\phi_i (\lambda)}{ \overline{F}(z)  c_i(\lambda)} -1.
\end{equation}
This requires estimates of  $ \overline{F}(z)$ and $ c_i(\lambda)$. In Sect.~\ref{sec:continuum}  we will discuss various methods of doing this.  In this section we use the true values of $ \overline{F}(z)$ and $c_i$, a procedure which, as we will see, only slightly overestimates the precision of the determination of the BAO signal. A grid with (12.8 Mpc$/h)^3$ cells was filled with the part of the $\delta_F$ 
spectra which corresponds to the \lya\ forest (we selected $1041<\lambda<1181\AA$, in the quasar rest frame). An unweighted average of $\delta_F$ was calculated for all pixels of all \lya\ forest spectra lying in a considered grid cell. Some grid cells do not contain any \lya\ forest spectra. The average of all other cells was computed and this average was subtracted from the content of all filled cells, while the content of unfilled cells remained zero. This was done in order to avoid including the Fourier transform of the quasar spatial distribution in the power spectrum. This is analogous to subtracting a synthetic catalog in the case of galaxy surveys, as advocated by \citet{1994ApJ...426...23F}. 
  
A Fourier transform of the resulting grid was performed and the modulus squared computed for each mode $(k_x,k_y,k_z)$, which gives the power for the considered mode. The angular-averaged power spectrum was then obtained as the average of the power for all the modes with $|k|$ in a given bin. Note, however, that we did not include purely radial $(0,0,k_z)$ and purely transverse $(k_x,k_y,0)$ modes. The former correspond to Fourier transforms along quasar lines of sight. These are severely affected by the uncertainty on the quasar unabsorbed spectrum estimate. The latter have a very large sampling term contribution due to the large value of $P_F^{1D}(k_\parallel=0)$. Purely transverse modes are 5.8\% of all modes at the first BAO peak and 3.4\% at the second BAO peak, while the number of purely longitudinal modes is less than 0.01\% of all modes at the first BAO peak. The error on $P(k)$ is obtained as the rms of the power in all modes in the considered $k$ bin, divided by the square root of the number of modes. The correlations between the power in different $k$ bins are expected to be small and we neglect them.   

\begin{figure} 
\resizebox{\hsize}{!}{
\includegraphics{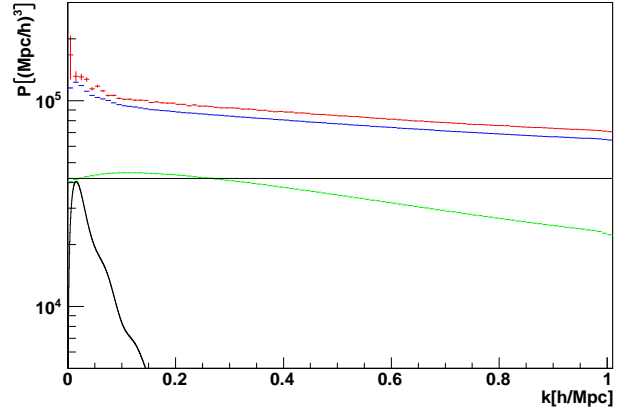}} 
\caption{Power spectrum in the simulation (red upper histogram) compared to the expectation of Eq.~\ref{Eq:Pk} (blue, just below red). The latter is the sum of the input power spectrum (black, decreasing quickly with $k$), the noise contribution (black, constant with $k$) and the sampling contribution (green, decreasing slowly with $k$).}
\label{fig:PkvsTh}
\end{figure}

The resulting power spectrum exceeds Eq.~\ref{Eq:Pk} by about 10\%, as illustrated in Fig.\ref{fig:PkvsTh}. We see that the sampling and noise contributions are of similar sizes and much larger than the input (LSS) power spectrum, which means that the BAO oscillations will be considerably diluted, as can be seen in Fig.~\ref{fig:PkP6}.

\subsection{Reconstruction of the BAO scale}
\label{sec:BAOscale}
 
\begin{figure} 
\resizebox{\hsize}{!}{
\includegraphics{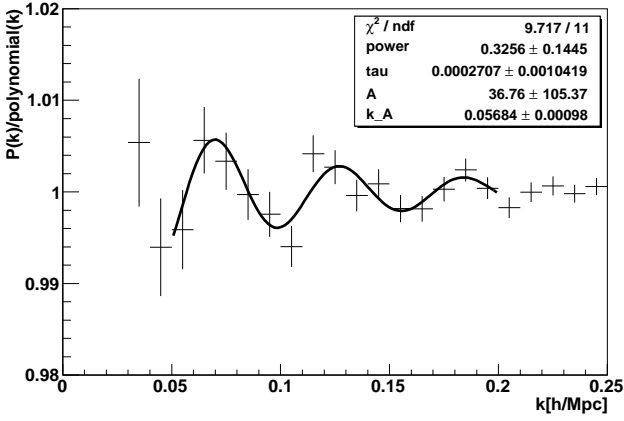}} 
\caption{Power spectrum for a realization of the nominal setup (15 QSO/deg$^2$) divided by a 7$^{th}$ order polynomial and fitted with \citet{2003ApJ...594..665B} formula, see text. }
\label{fig:PkP6}
\end{figure}

To infer the BAO scale, the angular-averaged power spectrum was first fitted by a polynomial. The power spectrum divided by this polynomial was then fitted with
\begin{equation}
\label{Eq:fit1D}
1+ A \left \{ k \exp \left[\left(\frac{-k}{\tau}\right)^p \right] \sin\left[2\pi \frac{k}{k_A}
\;\right] \right\}\; ,
\end{equation}
 as suggested by \citet{2003ApJ...594..665B}.  
This provides the (isotropic) BAO scale $k_A$, as illustrated in Fig.~\ref{fig:PkP6}. This figure corresponds to an average realization with $\Delta\chi^2=16.8$ and the BAO oscillations are quite visible. Their amplitude is less than a percent, which means they are considerably diluted relative to the input power spectrum where they are on the order of 10\%. This is due to the sampling and noise contributions to the observed power spectrum, see Eq.~\ref{Eq:Pk}. 

The upper limit of the polynomial fit range was set to 0.24, due to our 12.8 Mpc/$h$ FFT cells which result in a maximum $k$ of 0.245 $h$/Mpc in $x$, $y$ or $z$ direction. The lower limit and the order of the polynomial were set so as to get a good $\chi^2$ to the power spectrum obtained for some additional simulations which did not include the BAO features. These simulations were generated using the power spectrum with baryons but without BAO by \citet{eisensteinal98}. We noted that using a polynomial of too low a degree could result in a significant bias in the reconstructed BAO scale, while on the other hand increasing the polynomial degree beyond what is needed to get a good fit could degrade the performance (i.e.~increase the rms of $k_A$) because the polynomial fit starts to fit out the BAO features. Depending on the studied scenario, the lower limit was set between 0.02 and 0.05 $h$/Mpc and the order of the polynomial was either 6 or 7.  The range for oscillation fitting was set to $0.05<k<0.2$ $h$/Mpc, somewhat narrower to avoid edge effects in the polynomial fit.  We anyway do not want to include in this fit high $k$ values for which non-linear effects are important.

Ten Gaussian random field simulations were produced. Each of them was used ten times, with different random seeds to generate the quasar positions. One hundred values of $k_A$ were then obtained and their rms provides an estimate of the statistical precision on the BAO scale. This was done for several different scenarios and the results are presented in Sect.~\ref{sec:resuTrueC}.
The mean value of the error on $k_A$ returned by the Minuit fitting package~\citep{James:1975dr} is found to be compatible with the rms of $k_A$, which confirms that our estimate of the error on $P(k)$ and the fact that we neglected correlations are both reasonable. 

\subsubsection{Radial and transverse BAO scales}

In addition to the anisotropy of $P(\vec k)$ due to peculiar velocities, the reconstructed power spectrum, $P_{\rm obs}(\vec k)$, involves a stronger anisotropy due to the sampling contribution in Eq.~\ref{Eq:Pk} which depends on $k_\parallel$. It is difficult to find a functional form to fit the 2D power spectrum $P(k_\perp,k_\parallel)$. Instead $P(k_\perp,k_\parallel)$ was divided by the polynomial obtained from the 1D fit (previous section) to provide a ``reduced" power spectrum, which was then smoothed using  algorithms implemented in the CERN ROOT package. The ratio of the reduced-power-spectrum to the smoothed  reduced-power-spectrum was fitted with 
\begin{equation}
\label{Eq:fit2D}
1+ A \left \{ k \exp \left[\left(\frac{-k}{\tau}\right)^p \right] \sin\left[2\pi 
\sqrt{\frac{k_\perp}{k^A_\perp} +\frac{k_\parallel}{k^A_\parallel}}\;\right] \right\}.
\end{equation}
The rms from the 100 resulting values of $k^A_\perp$ and $k^A_\parallel$ provides an estimate of the error on the transverse and radial BAO scales.
Note however that this 2D fitting is quite delicate. It does not work in all cases and a more sophisticated procedure would be needed to analyze real data.  

\section{Results}
\label{sec:results}

\subsection{Results with the true quasar unabsorbed spectrum}
\label{sec:resuTrueC}

In this section we discuss the performance of BAO reconstruction in the ideal case where in Eq.~\ref{Eq:delta_F} we divided by the true $\overline{F}(z)$ and $c_i(\lambda)$. By default the simulation was performed for 15 QSO/deg$^2$ in the redshift range $2.2<z<3.5$ and with magnitude $18<m_g<22$. The results were slightly scaled to correspond to the 10,000 deg$^2$ of the BOSS survey. Different scenarios in terms e.g.~of number of quasars per deg$^2$ or noise level were studied. Table~\ref{tab:resu} presents, for each scenario, the rms in percent for the isotropic ($k_A$), radial ($k_\parallel$), and transverse ($k_\perp$) BAO scales.
To assess the significance of the BAO detection, we also give the difference of $\chi^2$ relative to the case with no BAO ($A=0$ in Eq.~\ref{Eq:fit1D}).

We start without noise and not including the effect of velocities (Table~\ref{tab:resu}, line 1). In this case, the transverse scale is much better reconstructed than the radial one because there are more transverse than radial modes (there are two transverse directions and only one radial).
Including peculiar velocities amplifies the power spectrum in the radial direction and dramatically improves the radial scale reconstruction, as can be seen on line 2.

\begin{figure} 
\resizebox{\hsize}{!}{
\includegraphics{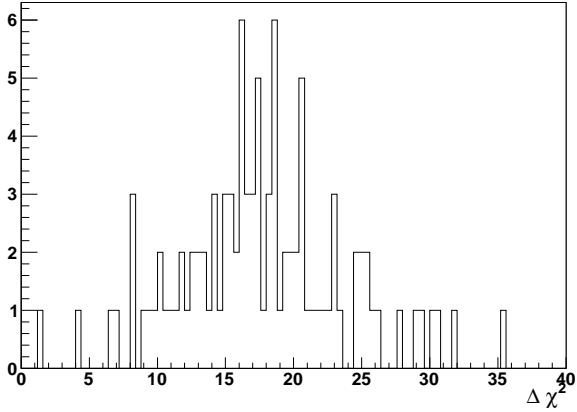}} 
\caption{Nominal simulation (15 QSO/deg$^2$). Difference between the $\chi^2$ for the nominal fit (Eq.~\ref{Eq:fit1D}) and the $\chi^2$ for no oscillations ($A=0$ in Eq.~\ref{Eq:fit1D}) for the 100 realizations.}
\label{fig:dChi2}
\end{figure}

The nominal case is obtained when we add noise according to BOSS setup, which results in rms of $2.28\pm0.16$,  $6.79\pm0.48$ and $3.86\pm0.27$ per-cent for the isotropic, transverse and radial BAO scales, respectively (line 4 of Table~\ref{tab:resu}).  On average over the 100 simulations, $\Delta\chi^2$ is 17.4, which means a significant detection of BAO features. This is, however, just an average and the difference of $\chi^2$ ranges from 2 to 35, as illustrated by Fig.~\ref{fig:dChi2}. 
At this point, we note that the (2Gpc$/h)^3$ Horizon simulation has a 17 times smaller angular coverage than our simulation. The $\chi^2$ differences for Horizon simulation would therefore be of order unity for BOSS \lya\ survey, so, as announced in Sect.~\ref{sec:Horizon}, the Horizon simulation is clearly not large enough to study the reconstruction of the BAO features with BOSS survey.

For the same nominal setup, an updated version (P.~McDonald, private comm.) of the analytic estimate of \citet{mcdoal07} gives an error of 1.8\% on the isotropic BAO scale when weighting pixels according to signal-to-noise ratio. Without weighting the computed error increases to 1.91\%. { Our result, $2.28\pm0.16$, is larger by a factor $1.19 \pm 0.08$, which is consistent with the fact that we find a power spectrum about 10\% larger than predicted by formula~\ref{Eq:Pk}, as illustrated in Fig.\ref{fig:PkvsTh}.
\cite{GreigBolton11} very recently published simulation results; they get an error of 1.38\% on the BAO scale for a fixed ratio $S/N=5$, whereas  we have in average $S/N=3.1$. When we decrease the $S/N$ ratio by a factor 1.7, we observe an increase of the rms by a factor 1.75, so it is not unlikely that an increase of $S/N$ by a factor 5/3.1=1.6, reduces the rms by a factor 2.28/1.38=1.65.

 }


{ Comparison of lines 3, 4 and 5 in Table~\ref{tab:resu} illustrates the variations of the performance with the quasar density. The sampling and noise contributions to the power spectrum (Eq.~\ref{Eq:KP}) scale as $1/N_{QSO}$. So, if we could completely neglect $P_F(k)$ in Eq.~\ref{Eq:KP}, we would expect the rms of the BAO to scale as $1/N_{QSO}$. We  observe a decrease of the rms with the quasar density, which is not as strong as $1/N_{QSO}$ but statistically compatible with the $N_{QSO}^{-0.61}$ dependence found by \cite{GreigBolton11}.} We investigated the dependence on the noise level.  Line 6 shows that increasing the noise by a factor 1.7 results in a significant degradation of the performance. As is clear from Fig.~\ref{fig:PkvsTh}, increasing the noise power spectrum by a factor $1.7^2$ makes it the dominant contribution to the total power spectrum.

We also considered the case of the BigBOSS project, assuming 60 QSO/deg$^2$ up to a $g$-magnitude of 23, over 14,000 deg$^2$ (last line of Table.~\ref{tab:resu}). With a 4 meter telescope and 5 exposures of 900 s instead of 2.5 m and 4 exposures, the noise is reduced by nearly a factor two relative to BOSS case. The resulting rms of the BAO scale is improved by a factor 4.2 relative to BOSS nominal case.


Finally, we note that naively combining the rms on the transverse and radial scales in Table~\ref{tab:resu} results in an error which is significantly larger than the rms on the isotropic scale. This is (partly) due to a significant  anticorrelation between the reconstructed transverse and radial scales. When the rms is quite small, e.g.~in the nominal case with 20 QSO/deg$^2$ or in the BigBOSS case, taking into account a correlation coefficient of typically $-0.40$, results in a combined rms in agreement with the rms on the isotropic case. In other cases, the combined rms is still larger by a factor on the order 1.2, up to a factor 1.4 for the case without velocities. This confirms that fitting the BAO features in two dimensions is difficult, in particular when the quasar density is low.

\begin{table}
\caption{Rms of the BAO scale and significance of BAO feature detection, for different scenarios. 
}
\label{tab:resu}
\centering 
\begin{tabular}{cccccccc} \hline \hline
QSO\tablefootmark{a}	& Area\tablefootmark{b} & Noise\tablefootmark{c} & Velo\tablefootmark{d}	
& $k_A$\tablefootmark{e} & $k_\perp$\tablefootmark{f} & $k_\parallel$\tablefootmark{g}
					& $\Delta\chi^2$\tablefootmark{h} \\ 
(deg$^{-2}$) & (deg$^2$) &  &  & (\%) & (\%) & (\%)	&						\\ \hline
15     & 10,000 & 0     & -     & 2.67  & 5.26  & 10.5   & 15.8	\\    
15     & 10,000 & 0     & +     & 1.60  & 5.77  & 2.84   & 34.2 	\\ \hline      
10     & 10,000 & 1     & +     & 3.13  & 10.7  & 5.27   & 10.7 	\\      
15     & 10,000 & 1     & +     & 2.28  & 6.79  & 3.86   & 17.4 	\\	
20     & 10,000 & 1     & +     & 2.20  & 6.33  & 3.20   & 24.9 	\\ \hline  
15     & 10,000 & 1.7   & +     & 3.99  & 10.8  & 6.00   & 12.4 	\\      
60     & 14,000 & 0.5   & +     & 0.54  & 1.28  & 0.84   & 284  	\\    
\hline
\end{tabular}
\tablefoot{
\tablefoottext{a}{assumed number of QSO/deg$^2$} 
\tablefoottext{b}{survey size} 
\tablefoottext{c}{scaling factor applied to the noise level, i.e.~0 means no noise and 1 noise corresponding to BOSS nominal setup.} 
\tablefoottext{d}{indicates whether the effect of peculiar velocities was taken into account or not. }
\tablefoottext{e}{rms for the isotropic BAO scale}
\tablefoottext{f}{transverse BAO scale}
\tablefoottext{g}{radial BAO scale}
\tablefoottext{h}{significance of BAO feature detection}\\
With 100 simulations, the statistical error on the rms, is just rms$/\sqrt{200}=0.071*$rms, e.g.~0.16, 0.40 and 0.28\% for the nominal case (line 4). 
}
\end{table}


\subsection{Effect of unabsorbed spectrum error}
\label{sec:continuum}

Using Eq.~\ref{Eq:delta_F} to obtain $\delta_F$ requires an estimate of $\overline F(z) \times c_i(\lambda)$ for each spectrum and the above results were obtained using the true value of this product. One can get a fair estimate of $\overline F(z)$ from the observed spectra, up to a normalization factor which is completely degenerate with the normalization of the $c_i$. On the other hand, with the resolution and the $S/N$ ratio of the BOSS survey, the unabsorbed spectra, $c_i(\lambda)$, cannot be accurately determined from the observed spectra (see Fig.~\ref{fig:spectrum}). A possible approach is to fit a power law in $\lambda$ on the red side of the \lya\ emission line, to extrapolate it in the \lya\ forest, possibly multiplying it by some average shape of the unabsorbed spectrum as a function of $\lambda_{rf}$, the quasar rest-frame wavelength \citep[e.g.][]{xipush}. We could not do that because our mocks are based on PCA by \citet{2005ApJ...618..592S} which extend only up to 1600 $\AA$ in the quasar rest frame and therefore do not allow for a reliable power law fit. Reliable fits will be possible with the PCA provided very recently by \citet{Paris11}, extending up to 2000 $\AA$.

Instead, we computed the average spectrum in the forest, as a function of $\lambda_{rf}$. Then we divided each spectrum by this average spectrum, fitted the result with a power law in $\lambda$, and finally divided by this power law. This is dividing the mock spectrum by an estimate of $\overline F(z) \times c_i(\lambda)$. For the nominal setup, this results in an rms of $(2.55 \pm 0.16)$\% for the BAO scale, only a factor $1.12\pm0.04$ larger than what is obtained with the true unabsorbed spectrum. We also note that, while we did not observe any bias on the reconstructed BAO scale when using the true unabsorbed spectrum, with this approximate unabsorbed spectrum, there is a bias of $-0.59\pm0.18$\%, which is, however, quite smaller than the $2.55$\% statistical error on the BAO scale. In this procedure, since we do a power law fit along the quasar spectra, we remove large scale power in the radial direction, so all modes with low value of $k_\parallel$  are strongly reduced and $P(\vec k)$ becomes very anisotropic. In this case, the smoothing used in the 2D fit procedure does not make sense. So we do not get results for the transverse and radial BAO scales, separately. A more sophisticated procedure would need to be developed.

Another possible approach is to use PCA to predict the unabsorbed spectrum in the forest from the flux redward of the \lya\ emission line.  However, we cannot use mock spectra generated with PCA to study how precisely PCA reconstruct the unabsorbed spectrum. Instead we used 78 observed unabsorbed spectra, $c_{obs}(\lambda)$,  and their corresponding PCA estimates, $c_{PCA}(\lambda)$ provided by \citet{Paris11}.
The observed unabsorbed spectra were manually estimated from high $S/N$ SDSS II spectra, while the PCA unabsorbed spectra were obtained for each spectrum by a PCA analysis of the 77 other spectra. For each of our spectra, with true unabsorbed spectrum $c_i(\lambda)$, we randomly selected spectrum $j$ within the 78 spectra from \citet{Paris11}, and used as an unabsorbed spectrum $c_i(\lambda) \times c_{PCA}^j(\lambda) / c_{obs}^j(\lambda)$. Since there is much less uncertainty on $\overline F(z)$ we used the true value of this function.
This results in an rms of ($2.73 \pm 0.19$)\% for the BAO scale, i.e.~a factor $1.20\pm0.11$ larger than what is obtained when using the true unabsorbed spectrum, and the bias is $+0.45\pm0.20$\%. 
In addition, most of the unabsorbed spectrum estimates of \citet{Paris11} are quite good while a few are grossly wrong. Therefore a significant improvement is to be expected if a criteria can be found to identify a-priori the spectra for which PCA will fail so that a different approach can be used for these few spectra.
Finally, note that the PCA unabsorbed spectrum estimates of \citet{Paris11} were obtained for high $S/N$ spectra, but PCA estimates might not be very sensitive to additional noise.

\section{Constraint on cosmological parameters}
\label{sec:cosmo}

The observation of the transverse and radial BAO scales provides a measurement of $d_T(z)/s$ and $d_H(z)/s$, respectively,
where $d_T(z)$ is the comoving angular distance, $d_H(z)$ the Hubble distance and $s$ the sonic horizon at decoupling.
For a flat $\Lambda$CDM universe $d_T$ and $d_H$ read

\begin{equation}
d_T = \int_{0}^z\frac{(c/H_0)\,dz}{E(z)}\; ,
\hspace*{5mm}
d_H = \frac{c/H_0}
{E(z)} \;,
\label{dTdH}
\end{equation}
where $E(z)=\sqrt{\Omega_\Lambda+\Omega_M(1+z)^3}$.

\begin{figure} 
\resizebox{\hsize}{!}{
\includegraphics{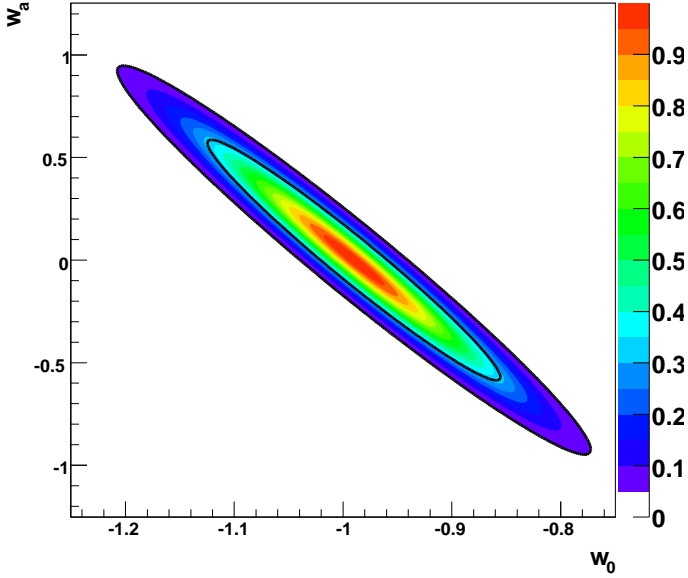}} 
\caption{Confidence level contours in the $(w_0,w_a)$ parameter plane
for Planck + BOSS LRG + BOSS \lya .}
\label{fig:w0wa}
\end{figure}


To estimate the sensitivity to parameters describing the dark energy equation of state, $p=w\rho$, we follow the procedure explained in \citet{2003ApJ...594..665B}. 
We introduce the $z$ dependence of $w$ as $w(z)=w_0 + w_a\cdot z/(1+z)$ and replace $\Omega_\Lambda$ in $E(z)$ in Eq.~\ref{dTdH}  by:
\begin{equation}
\Omega_\Lambda \longrightarrow \Omega_{\Lambda} \exp \left[ 3  \int_0^z   \frac{1+w(z^\prime)}{1+z^\prime } dz^\prime  \right]\;.
\end{equation}

Using the relative errors on the transverse (6.79\%) and radial (3.86\%) BAO scales, obtained for the nominal setup (Table~\ref{tab:resu}, line 4), and taking into account the anticorrelation, we can compute the Fisher matrix for  the five cosmological parameters $\Omega_m$, $\Omega_b$, $h$, $w_0$ and $w_a$.  We also use the Fisher matrix for Planck mission computed for the Euclid proposal \citep{Euclide09} which assumes a flat universe and involves the 8 parameters: $\Omega_m$, $\Omega_b$, $h$, $w_0$, $w_a$, $\sigma_8$, $n_s$ (spectral index of the primordial power spectrum) and $\tau$  (optical depth to the last-scatter surface). Combining BOSS and Planck Fisher matrices allows us to compute the errors on dark energy parameters. If we define the factor of merit as the inverse of the 1-$\sigma$ uncertainty ellipse in the $(w_0,w_a)$ plane, we get 34 for Planck and BOSS LRG survey. When we add BOSS \lya\ survey this increases to 48. The corresponding confidence level contours are plotted in Fig.~\ref{fig:w0wa}.


\section{Summary and conclusion}
\label{sec:discussion}

In this paper we have investigated the possibility to constrain the BAO scale from the quasar \lya\ forest and in particular the precision that will be reached by a survey such
as BOSS. To this aim, we have simulated realistic mock quasar spectra mimicking the survey. The volume of the largest N-body simulations being too small for such a study, we resorted to Gaussian random fields, combined with lognormal approximation and FGPA. We investigate the effect of noise, peculiar velocity and random unabsorbed quasar spectra generated using a principal component analysis (PCA).

The power spectrum of the transmitted flux fraction in the \lya\ forest was thus computed and was either fitted in two dimensions to reconstruct the radial and transverse BAO scales, or averaged over angles and fitted in one dimension to reconstruct the isotropic BAO scale. This was done over 100 realizations, resulting in an rms of 2.3\% on the isotropic BAO scale, or 6.8\% and 3.9\% on the transverse and radial BAO scales determinations separately. This is compatible with analytical estimates by \citet{mcdoal07}. The BAO features are detected with an average $\Delta\chi^2$ of 17 and the FOM for dark energy parameter determination improves from 34 for Planck and BOSS LRG survey to 48 when the \lya\ survey constraint is included. We note, however, that $\Delta\chi^2$ varies significantly between realizations, ranging from 2 to 35.

These above estimates were obtained assuming a perfect knowledge of the quasar unabsorbed spectrum. Errors on the estimate of the unabsorbed spectrum  will increase the errors on the BAO scale by 10 to 20\% and result in sub percent biases, overall quite small compared to the statistical error. Note that the effect of the presence of damped \lya\ and metal lines on the BAO measurement was not included in our mocks.

\begin{acknowledgements} We would like to thank Rupert Croft, Andreu Font-Ribera, Patrick McDonald and Anze Slosar for useful discussions and the last two for comments to the paper.
\end{acknowledgements}

\bibliographystyle{aa}	
\bibliography{bao}		

\end{document}